\documentclass[12pt,preprint]{aastex}

 \newcommand{\Li}{$^{8}$Li($\alpha$,n)$^{11}$B} \newcommand{\aansequence}{$^{4}$He($\alpha$n,$\gamma$)$^{9}$Be($\alpha$,n)$^{12}$C} \newcommand{\aan}{$^{4}$He($\alpha$n,$\gamma$)$^{9}$Be} \newcommand{\atg}{$^{4}$He(t,$\gamma$)$^{7}$Li} \newcommand{\be}{$^{9}$Be($\alpha$,n)$^{12}$C} \slugcomment{Submitted to Astrophysical J.} 
\begin{document}
%\title{A New Implication on dynamical $s/k$-$\tau$-$Y_{e}$ relation required to %successful r-process by considering \Li} 
\title{Light-Element Reaction flow and the Conditions for r-Process Nucleosynthesis} 

\author{T. Sasaqui\altaffilmark{1,2,5}}
\affil{National Astronomical Observatory of Japan} \affil{Department of Astronomy, Graduate School of Science, University of Tokyo}
\email{sasaqui@th.nao.ac.jp}

\author{K. Otsuki\altaffilmark{3}}
\affil{Department of Astronomy and Astrophysics, University of
Chicago}

\author{T. Kajino\altaffilmark{1,2}}
\affil{National Astronomical Observatory of Japan} \affil{Department of Astronomy, Graduate School of Science, University of Tokyo}

\and
\author{G. J. Mathews\altaffilmark{4}}
\affil{Center for Astrophysics,
Department of Physics, University of Notre Dame}

\altaffiltext{1}{National Astronomical Observatory of Japan 
2-21-1 Osawa, Mitaka, Tokyo,
181-8588, JAPAN}

\altaffiltext{2}{Department of Astronomy, Graduate School of Science, University of Tokyo, 7-3-1 Hongo, Bunkyo-ku, Tokyo 113-0033, Japan} 

\altaffiltext{3}{Department of Astronomy and Astrophysics, LASR103, University of
Chicago, 5640 South Ellis Avenue, Chicago, IL 60637, U.S.A.} 

\altaffiltext{4}{Center for Astrophysics, Department of Physics,
University of Notre Dame, Notre Dame,
IN 46556, U.S.A.}

\altaffiltext{5}{present address : Naka Division, Nanotechnology Products Business Group, Hitachi High-Technologies Corporation, 882, Ichige, Hitachinaka, Ibaraki, 312-0033, Japan}

\begin{abstract}
We deduce new constraints on the entropy per baryon ($s/k$), dynamical timescale ($\tau_{dyn}$), and electron fraction ($Y_{e}$) consistent with heavy element nucleosynthesis in the r-process. We show that the previously neglected reaction flow through the reaction sequence \atg (n,$\gamma$)\Li~ significantly 
enhances the production of seed nuclei. We analyze the r-process nucleosynthesis in the context of a schematic exponential wind model. We show that fewer neutrons per seed nucleus implies that the entropy per baryon required for successful r-process nucleosynthesis must be more than a factor of two higher than previous estimates. This places new constraints on dynamical models for the r-process. \end{abstract}


\section{Introduction}

The astronomical site for r-process nucleosynthesis by rapid neutron capture has not yet
been unambiguously determined.
Observations of metal poor stars
(e.g. Snedden et al. 1996) indicate an abundance pattern for the early Galactic r-process elements which is very similar to that of the Solar r-process abundance distribution. Hence, it is often argued that core-collapse supernovae ($e.g.$~Type II SNe) are the most likely site for r-process nucleosynthesis. Such events are the first contributors to the abundances observed at lowest metallicity. The possibility remains, however, that the r-process could be associated with neutron star mergers \citep{freiburghaus99} or gamma-ray burst environments \citep{inoue02} in which the required neutron-rich conditions can also be realized.

Moreover, the environment suitable for the r-process is not yet fully understood numerically.
Even in the presently popular paradigm of neutrino-driven winds from Type II SNe, physical conditions of the r-process environment are largely dependent on the details of the adopted numerical simulations \citep{meyer92, woosley94, witti94, takahashi94, qian96, cardall97, hoffman97, 
otsuki03, sumiyoshi00, wanajo01, otsuki03}.
% Many models could not help depending on the parametric search on s, $\tau$, and $Y_{e}$. 

As a useful guide for numerical studies of the r-process environment, \citet{hoffman97} determined empirical conditions required to produce the platinum peak in the r-process. They deduced a phenomenological 
constraint on the
parameter space of $s/k$, $\tau_{dyn}$, and $Y_{e}$, i.e. $ (s/k \propto Y_{e}{\tau_{dyn}}^{1/3})$.

In the present work we reinvestigate these phenomenological constraints and deduce a new allowed parameter space for $s/k$-$\tau_{dyn}$-$Y_{e}$. We deduce 
significantly greater restrictions on the r-process environment.
The difference of the present work with the previous study 
can be traced to the treatment of the
$\alpha$-process.
In formulating the $s/k$-$\tau_{dyn}$-$Y_{e}$ relation, Hoffman et al.~(1997) considered only the reaction flow through $^{4}$He($\alpha$n, $\gamma$)$^{9}$Be($\alpha$,n)$^{12}$C as an 
$\alpha$-process path (see section 4 in \citet{hoffman97}). In the present work, however, we also include the reaction sequence
$^{4}$He(t, $\gamma$)$^{7}$Li(n,$\gamma$)$^{8}$Li($\alpha$,n)$^{11}$B. 

Although the reaction sequence \aansequence ~is usually the dominant flow in the $\alpha$-process, 
alternative reaction sequences such as the \atg (n,$\gamma$)\Li path
%, $^{4}$He(t,$\gamma$)$^{7}$Li(n,$\gamma$) 
can provide enhanced reaction flow toward seed nuclei \citep{terasawa01}. The production of heavy nuclei is, therefore, quite sensitive
\citep{sasaqui05a,sasaqui05b} to the rates for these reactions. Also, since the work of Hoffman et al.~(1997), new measured rates
% \citet{boyd92},\ Gu et al. 1995,\ \citep{mizoi00}, \citep{paradellis90} % and
\citep{hashimoto04} are available for the \Li~ reaction. 

The purpose of this short paper is therefore to reformulate the $s/k$-$\tau_{dyn}$-$Y_{e}$ constraints on the SN dynamics by incorporating the important
effects from the reaction sequence
$^{4}$He(t,$\gamma$)$^{7}$Li(n,$\gamma$)$^{8}$Li($\alpha$,n)$^{11}$B.

%**** reaction speed ni tuite
\section{Calculation}
\subsection{Exponential Model}
For the present studies we utilize a schematic exponential model similar to that adopted by \citet{meyer92,meyer97}. This model provides an adequate approximation to the evolution of ejected material in a wide variety of the plausible conditions of the r-process such as may occur for example in both delayed and prompt SNe \citep{hillebrandt84, sumiyoshi01, wanajo03}, neutron-star mergers \citep{freiburghaus99}, or gamma-ray burst (GRB) environments. 

In this model the dynamical expansion timescale, $\tau_{dyn}$, denotes how rapidly the temperature evolves, \begin{eqnarray}
\tau_{dyn}^{-1}=-{{1}\over{T-T_{a}}}{{{\rm d}T}\over{{\rm d}t}} \label{tau_dyn_def},
\end{eqnarray}
where $T_a$ is the asymptotic temperature \citep{otsuki03} of the material. This temperature
determines the freeze-out of the neutron-capture flow. The model assumes adiabatic expansion. Hence, the entropy per baryon $s/k \propto {{T^{3}}/{\rho}} =$ constant.
The temperature and density thus evolve according to: \begin{eqnarray}
T_{9}(t)=T_{9}(0)\exp(-t/{\tau}_{dyn})+T_{a}~~, \label{ondo}\\
\rho(t) = \rho(0)\biggl(\frac{T_{9}(t)}{T_{9}(0)}\biggr)^3~~, \label{mitudo}
\end{eqnarray}
where we adopt $T_{9}(0)=8.40$, $T_{a}=0.62$, and $ \rho(0)=3.3\times 10^8$ g cm$^{-3}$ from Otsuki et al.~(2003) and Sasaqui et al.~(2005a).

\section{Nucleosynthesis Network}
We employ the nucleosynthesis reaction
network used in \citet{otsuki03}, which was derived from the 
network code described in
\citet{meyer92} and
\citet{woosley94}, and expanded in \citet{terasawa01}. Several further important modifications have also been made in the present reaction network. The main features are the following. 

The reaction \aan ~is still important even in wind models with a short dynamical expansion timescale. The three-body reaction rate for \aan
%$\alpha$($\alpha$n,$\gamma$)$^9$Be
~is taken from the network
estimate  of \citet{sumiyoshi02} based on recent experimental data \citep{utsunomiya01} for this reaction cross section which spans the low energy region of astrophysical interest. However, it now shares the main
nuclear reaction chain with a new flow path $^{4}$He(t,$\gamma$)$^{7}$Li(n,$\gamma$)$^{8}$Li($\alpha$,n)$^{11}$B \citep{terasawa01}.
The $^{8}$Li($\alpha$,n)$^{11}$B
reaction in particular has been identified (Terasawa et al. 2002; Sasaqui et al. 2004) \citep{terasawa02,sasaqui05a,sasaqui05b} to be critical in the production of
intermediate-to-heavy mass elements.
%in some inhomogeneous Big-Bang
%nucleosynthesis models (Malaney \& Fowler 1998,\ Boyd \& Kajino 1989,\ %Kajino \& Boyd 1990) as well as for r-process nucleosynthesis (Terasawa %et al. 2001,\ Kajino, Wanajo \& Mathews 2002). 
\citet{hashimoto04} have
carried out very precise measurements of the exclusive 
(i.e. individual states) reaction cross section 
%for $^{8}$Li($\alpha$,n)$^{11}$B$^{\ast}$ as well as 
for $^{8}$Li($\alpha$,n)$^{11}$B.
%$_{gs}$
Their results confirm
that transitions leading to several excited states of $^{11}$B make the
predominant contribution to the total reaction cross section. This is in
good agreement with the previous measurements of the inclusive (i.e.~sum
of excited states) reaction cross section \citep{boyd92,gu95,mizoi00}. %(\citet{boyd92},\ Gu et al.~(1995),\
%\citet{mizoi00})
Hence, we employ the newest
cross section data from \citet{hashimoto04}. 

\begin{figure}
\centering
%\rotatebox{0}{\includegraphics[width=18cm,height=15cm]{f5_ver5.eps}} 
\rotatebox{0}{\includegraphics[width=18cm,height=15cm]{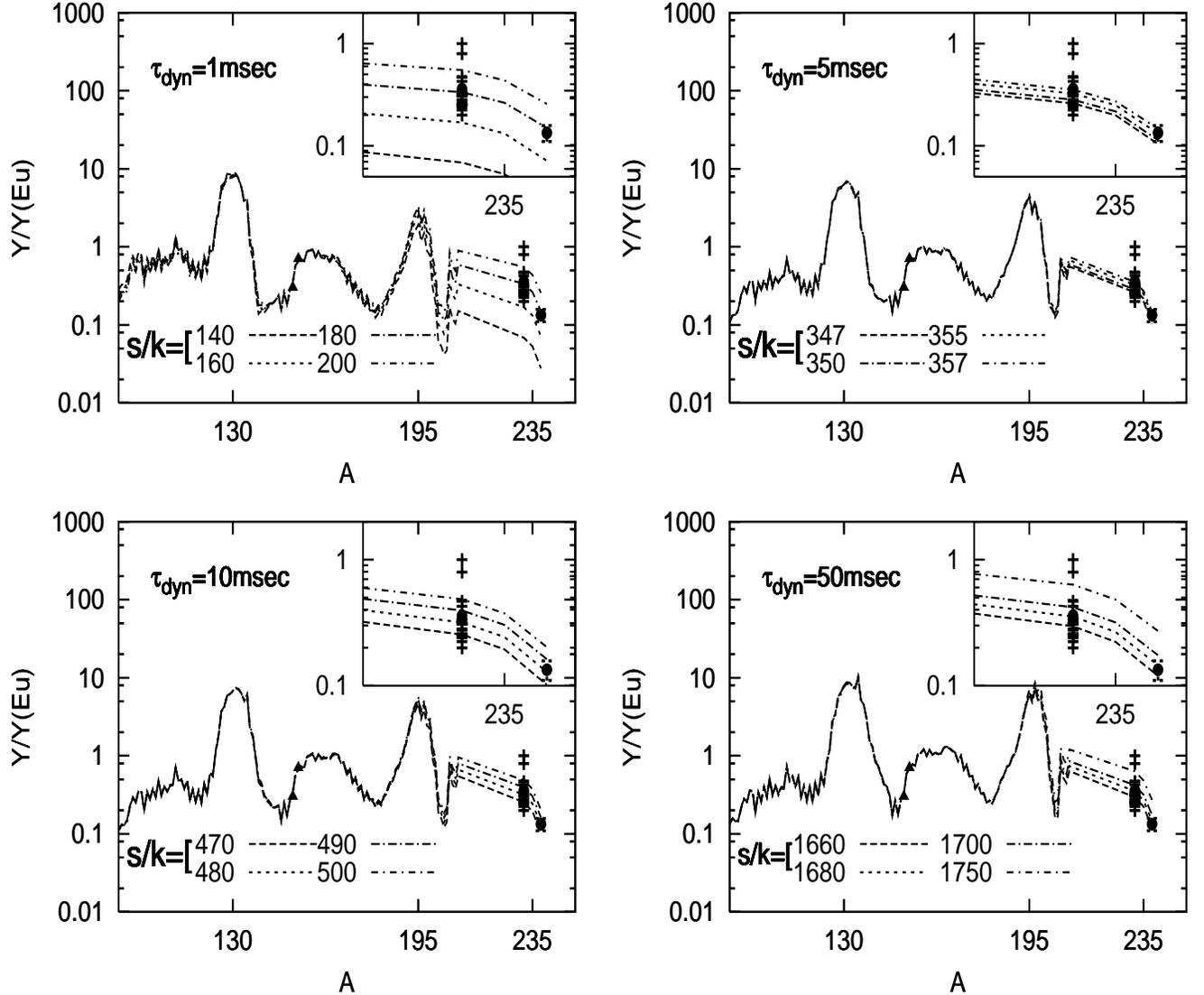}} 
%\rotatebox{0}{\includegraphics[width=18cm,height=15cm]{f1.eps}} 
\caption{Comparison of the final computed r-process abundances 
based upon our adopted  network \citep{otsuki03} for exponential models  with
$Y_e = 0.45$ and various values of $\tau_{dyn}$ and $s/k$ as labeled. 
Abundances are normalized to Y/Y(Eu) = 1 [for Y(Eu) = Y($^{151}$Eu) + Y($^{153}$Eu)].
For each figure $\tau_{dyn}$ is given in the upper left corner while values for $s/k$ are shown near the bottom. These models have been chosen so as to produce final abundances consistent with observation.  Inserts in the upper right corner show an expanded view of the calclated and observed Th and U abundances
as tabulated in Table 3 of  \citet{sasaqui05b}.
}
\label{f5}
\end{figure}

We also note that we calculate the nucleosynthesis sequence from nuclear statistical equilibrium (NSE), the $\alpha$-process, $\alpha$-rich freeze-out, the r-process and 
subsequent beta-decay and alpha-decay in a single network code rather than to split the calculation into two parts as was done in \citet{woosley94}.
It is important to compute self-consistently the evolution of seed nuclei along with heavy element production in the r-process \citep{sasaqui05a,sasaqui05b}. Computed final r-process abundances 
for $Y_e = 0.45$ and various values for the dynamical
timescale and entropy are shown in Figure \ref{f5}.  
As noted in Hoffman et al.~(1997),  this entire line of inquiry will only be relevant if the conditions 
important to seed production prior to the r-process occur in an environment with a neutron excess.

\section{Analytic Treatment Of The $\alpha$-process} 

As in \citet{hoffman97}, we analyze the $\alpha$-process in detail in order to provide new dynamical constraints on the $s/k$-$\tau_{dyn}$-$Y_{e}$ parameter space relevant to the r-process. The $\alpha$-process is particularly important as it is the means for producing seed nuclei for subsequent r-process neutron capture.

As the
temperature drops below $T_{9}\sim$5.0 the reaction flow falls out of NSE and 
the $\alpha$-process operates
until the temperature drops below $T_{9}\sim$2.5. During this process, $\alpha$ particles are consumed through the main bottleneck reaction sequence \aansequence ~and also the secondary reaction path \atg (n,$\gamma$)\Li. ~Seed nuclei for the r-process are subsequently produced by a sequence of $\alpha$-capture reactions starting with $^{9}$Be($\alpha$,n)$^{12}$C or \Li.

\subsection{$^4$He($\alpha$n,$\gamma$)$^{9}$Be($\alpha$,n)$^{12}$C } 
\label{beseq}

The approximate time evolution of the abundances of $\alpha$ particles ($Y_{\alpha}$) and neutrons ($Y_{n}$) is expressed as in Hoffman et al.~(1997), 
\begin{eqnarray}
{{\rm d Y_{\alpha}}\over{\rm d t}} \approx -{{\bar{Z}}\over{2}} Y_{\alpha} Y_{9} \rho N_{A} \langle\sigma v\rangle_{\alpha n},
\label{11a}
\end{eqnarray}
\begin{eqnarray}
{{\rm d Y_{n}}\over{\rm d t}} \approx  -(\bar{A}-2\bar{Z}) Y_{\alpha} Y_{9} \rho N_{A} \langle\sigma v\rangle_{\alpha n},
\label{11b}
\end{eqnarray}
where $Y_{9}$ is the abundance of $^{9}$Be and $N_{A}\langle\sigma v\rangle_{\alpha n}$ is the $^{9}$Be($\alpha$,n)$^{12}$C reaction rate .  The  quantities $\bar{A}$ and $\bar{Z}$ are the mean mass number and mean proton number, respectively, of typical seed nuclei as defined in \citet{hoffman97}.

Because of the low $Q$-value for the \be ~reaction rate, statistical equilibrium is realized between $^{9}$Be and $^{4}$He over the temperature range of interest. Hence, we can write \citep{hoffman97}
\begin{eqnarray}
Y_{9}= Y(4,9) \approx G(4,9)[\zeta(3)^8 \pi ^{-4} 2^{11}] 9^{3/2} ({{kT}\over{m_{N}c^2}})^{12}
\phi ^{-8} Y_{p}^{4} Y_{n}^{5}
\exp{({{B(4,9)}\over{kT}})}, \nonumber
\end{eqnarray}
\begin{eqnarray}
Y_{\alpha}=Y(2,4) \approx G(2,4)[\zeta(3)^3 \pi ^{-3/2} 2^{7/2}] 4^{3/2} ({{kT}\over{m_{N}c^2}})^{9/2}
\phi ^{-3} Y_{p}^{2} Y_{n}^{2}
\exp{({{B(2,4)}\over{kT}})}~~. \nonumber \end{eqnarray}
Here,
B(4,9)=58.16 MeV and B(2,4)=28.29 MeV. Therefore, \begin{eqnarray}
Y_{9} \approx 8.66 \times 10^{-11} \rho_{5}^{2} T_{9}^{-3} Y_{\alpha}^2 Y_{n} \exp{({{18.26}\over{T_{9}}})}~~. \nonumber \end{eqnarray}
Adopting the exponential dynamical model [i.e.~Eqs.~(\ref{ondo}) and (\ref{mitudo})],
equations (\ref{11a}) and (\ref{11b}) become, \begin{eqnarray}
{{\rm d Y_{\alpha}}\over{\rm d T_{9}}} \approx  {{\bar{Z}}\over{2}} Y_{\alpha}^3 Y_{n} f(T_9)\tau_{dyn}, \label{13a}
\end{eqnarray}
\begin{eqnarray}
{{\rm d Y_{n}}\over{\rm d T_{9}}} \approx  (\bar{A}-2\bar{Z}) Y_{\alpha}^3 Y_{n} f(T_9)\tau_{dyn}, 
\label{13b}
\end{eqnarray}
where $f(T_{9})$ is given by
\begin{eqnarray}
f(T_9) \approx 8.66 \times 10^{-6}\rho_{5}^3 T_{9}^{-4} \exp{(18.31/T_{9})}N_{A}\langle\sigma v\rangle_{\alpha n}~~{\rm sec}^{-1}~~. 
\end{eqnarray}
Now inserting $\rho_{5} \sim 3.33 {T_9}^3/(s/k)$, we have
\begin{eqnarray}
f(T_{9})\approx 3.20 \times 10^{-4}({s/k})^{-3} T_{9}^{5} \exp{(18.31/T_{9})}N_{A}\langle\sigma v\rangle_{\alpha n} ~~{\rm sec}^{-1}~~. 
\end{eqnarray}
Now integrating Eq.~(\ref{13b}) in the range between $T_{9}=2.5$ and $T_{9}=5.0$ where the $\alpha$-process is dominant, we obtain the final neutron abundance,
 \begin{eqnarray}
Y_{n,f} \approx Y_{n,0} \exp{\big[- (\bar{A}-2\bar{Z}) Y_{\alpha,0}^3 \tau_{dyn} \int_{2.5}^{5.0}f(T_9){\rm d}T_{9} \big]},
 \end{eqnarray}
where we have made use of the fact that \(Y_{\alpha} \approx Y_{\alpha,0}=X_{\alpha,0}/4 \approx 1/2Y_{e,i}\) during the $\alpha$-process. 

The integral can be approximated by \(\int_{2.5}^{5.0}{f(T_9){\rm d}T_{9} } \approx 7.44 \times 10^{8} {(s/k})^{-3}\). We introduce \citep{hoffman97} % and
%$N_{A}<\sigma v>_{\alpha n}$ is given by ****Q1 $Y_{n,f}=X_{n,f}\approx (({1-\bar{A}/A})/({1-2\bar{Z}/A}))(1-2Y_{e,i})$, $Y_{n,0}=X_{n,0} \approx 1-2Y_{e,i}$ and $Y_{\alpha} \sim 1/2 Y_{e,i}$, where $A$ is the mass number of
the produced r-process nucleus of interest. This leads to a lower limit on the entropy per baryon required to produce an r-process nucleus of mass number $A$, 
\begin{equation}
s/k \approx Y_{e,i} \biggl\{ {{9.3\times 10^{7}(\bar{A}-2\bar{Z})}\over{\ln{\left[(1-2\bar{Z}/A)/(1-\bar{A}/A)\right]}}} \tau_{dyn} \biggr\}^{1/3}~~. 
\label{result1}
\end{equation}
The results of this analysis is  expressed in the lower part of Figure \ref{f1-2-3}.  These expressions  are essentially identical to those of \citet{hoffman97}.  The
only differences stem from the use a different reaction rate in the
\aansequence ~sequence.  The linear scaling of this result on the choice of $Y_e$ is also apparent.

\subsection{$^4$He(t,$\gamma$)$^7$Li(n,$\gamma$)\Li} 
\label{liseq}

As described in Section 3, the reaction flow through \atg (n,$\gamma$)\Li ~is also important. We make an analogous treatment of
the \Li~reaction to that of the \be ~reaction sequence of Section \ref{beseq}. Hence, we write,
\begin{eqnarray}
{{\rm d Y_{\alpha}}\over{\rm d t}} = -{{\bar{Z}}\over{2}} Y_{\alpha} Y_{8} \rho N_{A} \langle\sigma v\rangle_{\alpha n},
\label{11c}
\end{eqnarray}
\begin{eqnarray}
{{\rm d Y_{n}}\over{\rm d t}} = -(\bar{A}-2\bar{Z}) Y_{\alpha} Y_{8} \rho N_{A} \langle\sigma v\rangle_{\alpha n},
\label{11d}
\end{eqnarray}
where $Y_{8}$ is the abundance of $^{8}$Li, and in this case $N_{A}\langle\sigma v\rangle_{\alpha n}$ is the reaction rate of \Li.

Because of the low $Q$-value for the \Li ~reaction, 
%rate is much faster than that  of \atg ~and$^{7}$Li(n,$\gamma$)$^{8}$Li,
statistic equilibrium is again realized 
\citep{hoffman97} between $^{9}$Li and $^{4}$He. 
Hence, we write 
\begin{eqnarray}
Y_{8}= Y(3,8)=G(3,8)[\zeta(3)^7 \pi ^{-7/2} 2^{19/2}] 8^{3/2} ({{kT}\over{m_{N}c^2}})^{21/2}
\phi ^{-7} Y_{p}^{3} Y_{n}^{5}
\exp{({{B(3,8)}\over{kT}})} ~~,
\end{eqnarray}
where, B(3,8)=41.28 MeV.
Using $Y_{\alpha}$ as defined in \S 4.1 then we deduce \begin{eqnarray}
Y_{8}\approx 7.96\times 10^{-14} Y_{n}^2 Y_{\alpha}^{3/2} \exp(-{{13.39}\over{T_9}})T_9^{-15/4}\rho_5^{5/2}. \end{eqnarray}
Once again
%\begin{eqnarray}
%Y_{9} \approx
%\end{eqnarray}
adopting an exponential model [Eqs.~(\ref{ondo}) and (\ref{mitudo})],
equation (\ref{11d}) becomes
%\begin{eqnarray}
%{{\rm d Y_{\alpha}}\over{\rm d T_{9}}} = {{\bar{Z}}\over{2}} Y_{\alpha}^3 Y_{n} g(T_9)\tau_{dyn}, %\label{13a}
%\end{eqnarray}
\begin{eqnarray}
{{\rm d Y_{n}}\over{\rm d T_{9}}} = (\bar{A}-2\bar{Z}) Y_{\alpha}^{3/2} Y_{n}^{2} h(T_9)\tau_{dyn}, \label{13d}
\end{eqnarray}
where $h(T_{9})$ is given by
\begin{eqnarray}
h(T_9) \approx 7.6 \times 10^{-9} {(s/k)}^{-7/2}T_{9}^{23/4}\exp{(-13.39/T_{9})}N_{A}\langle\sigma v\rangle_{\alpha n}~~. \end{eqnarray}
Here we have made use of the fact that \(s \sim 3.33{{{T_9}^3}/{\rho_{5}}}\). 
%\begin{eqnarray}
%\approx 3.20 \times 10^{-4}S^{-3} T_{9}^{-4} \exp{(18.31/T_{9})}N_{A}<\sigma v>. 
%\end{eqnarray}
Integrating Eq.~(\ref{13d}) from $T_{9}=2.5$ and $T_{9}=5.0$ we have,
\begin{eqnarray}
-{1\over{Y_{n,f}}}+{1\over{Y_{n,0}}} = (\bar{A}-2\bar{Z})\tau_{dyn} Y_{\alpha, 0}^{3/2} \int_{2.5}^{5.0}h(T_9) {\rm d}T_{9}, \nonumber
\end{eqnarray}
where we again use the fact that \(Y_{\alpha} \approx Y_{\alpha,0}=X_{\alpha,0}/4 \approx 1/2Y_{e,i}\) during the $\alpha$-process and we invoke the approximation, \(\int_{2.5}^{5.0}h(T_9){\rm d}T_{9} \approx 3.6 \times 10^{3} s^{7/2}\) for
$N_{A}\langle\sigma v\rangle_{\alpha n}$ given by \citet{hashimoto04}.

As a result,
\begin{equation}
s/k \approx \biggl\{{{3.57 \times 10^{3} 
Y_{e,i}^{3/2}(\bar{A}-2\bar{Z})(1-\bar{A}/A)(1-2Y_{e,i})}\over{2^{5/2}\bar{Z}/A}} \tau_{dyn} \biggr\}^{2/7}.
\label{result2}
\end{equation}
This result is also shown on the lower part of Figure \ref{f1-2-3}.  Here, the need that $Y_e < 0.5$ 
is evident as is  the scaling of these results with $Y_e$.

\subsection{Total Sequence}
Combining these two reaction branches,
the total change of neutron density with temperature now becomes:
\begin{eqnarray}
\biggl( \frac {\rm d Y_{n}} {\rm d T_{9}} \biggr)_{tot} = (\bar{A}-2\bar{Z}) \biggl[Y_{\alpha}^{3} Y_{n} f(T_9) + Y_{\alpha}^{3/2} Y_{n}^{2} h(T_9)\biggr] \tau_{dyn}, \label{13e}
\end{eqnarray}

Integrating Eq.~(\ref{13e}) from $T_{9}=2.5$ to $T_{9}=5.0$ we have,
\begin{eqnarray}
{1\over{Y_{n,f}}}&=&{1\over{Y_{n,0}}}\exp{\Biggl[ \int_{2.5}^{5.0}f(T_9)(\bar{A}-2\bar{Z})\tau_{dyn}Y_{\alpha}^{3}{\rm d}T_{9} \Biggr]}\\ \nonumber &&
+\int_{2.5}^{5.0}(\bar{A}-2\bar{Z})\tau_{dyn}Y_{\alpha}^{3/2}h(T_{9}) \exp{\biggl[(\bar{A}-2\bar{Z})\tau_{dyn}Y_{\alpha}^3\int_{2.5}^{T_{9}}f(T_{9}^{'}){\rm d}T_{9}^{'}}\biggr] {\rm d}T_{9},
\label{sol_tot}
\end{eqnarray}
[See Appendix B, Eq.~(\ref{ge-sol}) for a derivation of this result]. This leads to a new lower limit on the entropy required to produce an r-process nucleus with mass number $A$. This relation can be simplified because it can be approximately separated into two components (see Appendix C). One of them is for the reaction sequence $^4$He($\alpha$n,$\gamma$)$^{9}$Be($\alpha$,n)$^{12}$C and the other is  for the $^4$He(t,$\gamma$)$^7$Li(n,$\gamma$)\Li ~sequence. This leads to the following expression which  satisfies the relations (\ref{result1}) and (\ref{result2}).
\begin{eqnarray}
{1\over{Y_{n,f}}} \approx \Biggl(
{{1}\over{Y_{n,0}}}+{{\alpha_1}\over{\beta_0}} (1-e^{-2.5\beta_0})Y_{n,f}h(5.0)
\Biggr)
\exp\biggl[
\alpha_0\int_{2.5}^{5.0}f(T_9){\rm d}T_9 \biggr].\nonumber
\label{approx1}
\end{eqnarray}
Here,
\(\alpha_{0} = (\bar{A}-2\bar{Z})\tau_{dyn}Y_{\alpha}^{3}\), \(\alpha_{1} = (\bar{A}-2\bar{Z})\tau_{dyn}Y_{\alpha}^{3/2}\), and \(\beta_0=\biggl({{h^{'}(T_i)}\over{h(T_i)}}+ \alpha_0 f(T_i)\biggr)\).
Since the left side \({1/{Y_{n,f}}}\) is fixed once the initial conditions are specified, it must be constant. Hence, the right hand side must be constant as well.
By this requirement, we suppose that the both parts of the right hand side are constant. The term in parentheses on the right hand side yields \(s/k \propto \tau_{dyn}^{2/7}\) (similar to the result of \S 4.2, i.e. Eq. (\ref{result1})).
The later exponential term on the right introduces \(s/k \propto \tau_{dyn}^{1/3}\) (similarly to Eq. (\ref{result2}) of \S 4.1 ). As a result we find:
\begin{equation}
s/k
\propto
\tau_{dyn}^{13/21}
%\approx
%4.69\times 10^{3} Y_{e}^{10/7} \nonumber \\ %\{{{(\bar{A}-2\bar{Z})}\over{\ln{\left[(1-2\bar{Z}/A)/(1-\bar{A}/A)\right]}}} \}^{1/3}\{{{ % (\bar{A}-2\bar{Z})(1-\bar{A}/A)(1-2Y_{e,i})}\over{2^{5/2}\bar{Z}/A}} % \}^{2/7} \tau_{dyn}^{13/21}
\label{result3}
\end{equation}
The exact solution cannot be expressed analytically like Eqs.~(\ref{result1}) or (\ref{result2}). However,
the details of the numerical calculation are also shown on Figure \ref{f1-2-3} and compared with the
above analytic
results. The analytic result is in good agreement with the numerical simulation. 

\section{Results and Summary}

As noted in \S 4, the reaction sequence \atg (n,$\gamma$)\Li ~can be a competitor 
to the \aansequence ~sequence in the $\alpha$-process.
We have analyzed this by adding the contribution of the reaction \Li ~to the previous analysis
\citep{hoffman97} of the entropy constraint  %Since the dynamical parameters s, $\tau_{dyn}$, and $Y_{e}$ satisfy %both relation eq. \ref{result1} and eq. \ref{result2}, we think s must satisfy %the production of right side of \ref{result1} and that of \ref{result2}. %Namely,
%\begin{eqnarray}
%s/k &\approx& Y_{e,i} \{{{9.3\times 10^{7}(\bar{A}-2\bar{Z})}\over{\ln{\left[(1-2\bar{Z}/A)/(1-\bar{A}/A)\right]}}}\tau_{dyn} \}^{1/3} \nonumber\\ %&&\times \{{{3.57 \times 10^{3}
% Y_{e,i}^{3/2}(\bar{A}-2\bar{Z})(1-\bar{A}/A)(1-2Y_{e,i})}\over{2^{5/2}\bar{Z}/A}} % \tau_{dyn} \}^{2/7} \nonumber\\
%&\approx&
%4.69\times 10^{3} Y_{e}^{10/7} \nonumber \\ %&&\times \{{{(\bar{A}-2\bar{Z})}\over{\ln{\left[(1-2\bar{Z}/A)/(1-\bar{A}/A)\right]}}} \}^{1/3}\{{{ 
%(\bar{A}-2\bar{Z})(1-\bar{A}/A)(1-2Y_{e,i})}\over{2^{5/2}\bar{Z}/A}} % \}^{2/7} \tau_{dyn}^{13/21}
%\label{result}
%\end{eqnarray}

\begin{figure}
\centering
\rotatebox{0}{\includegraphics[width=18cm,height=15cm]{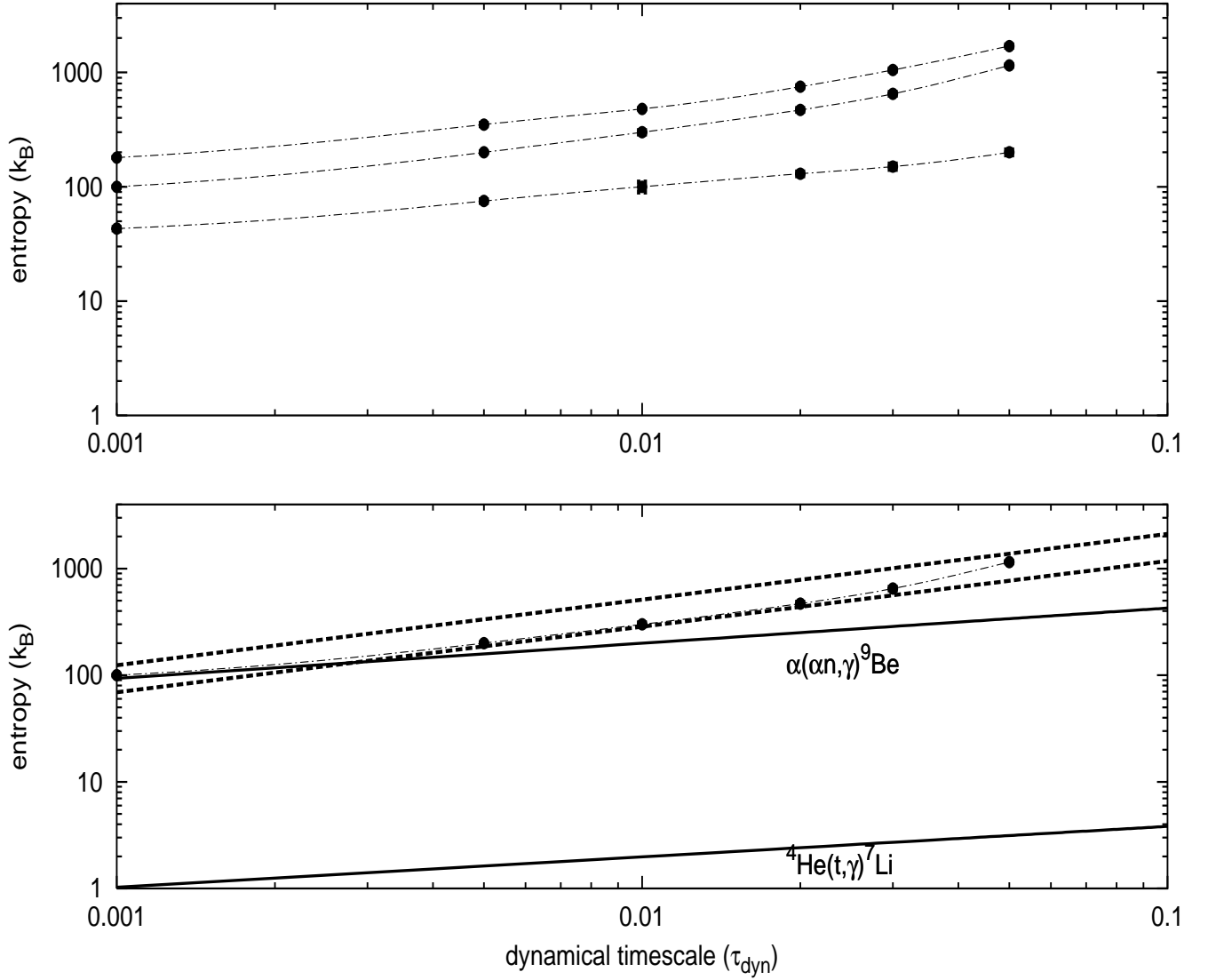}} 
%\rotatebox{0}{\includegraphics[width=18cm,height=15cm]{f2.eps}} 
\caption{The minimum entropy required as a function of the dynamical timescale
to produce the observed abundance of  characteristic  nuclei in the r-process based upon the numerical network calculation of Figure \ref{f5}.  The upper figure (points and dot-dashed lines) shows the entropy required to produce the 1st (lower curve), 2nd (middle curve)  r-process peaks and (A=232) actinide nuclei (upper curve).  The numerical result for the 2nd r-process peak is also shown on the lower figure where it is compared with the
the analytic limits (dotted lines) deduced for the 2nd (lower line) r-process peak and (A=232) actinide nuclei (upper line).
The solid lines show the entropy required when considering only the \aan ~flow of Eq.~(\ref{result1}) ($i.e.$ \citet{hoffman97}) or the \atg (n,$\gamma$)\Li  ~flow alone as labeled. }
\label{f1-2-3}
\end{figure}

In Figure
%\ref{f1}, \ref{f2}, and \ref{f4}).
\ref{f1-2-3} we compare this analytic model for the entropy constraint as a function of dynamical timescale with the nuclear simulation results . We have selected a comparatively wide range model parameters, $28 < \bar{Z} < 36$, and $85 < \bar{A} < 105$  after \citet{hoffman97}. In the present 
analysis we also consider 
 the production of the actinide nuclei ($^{232}$Th, $^{235}$U, and $^{238}$U).
 We consider such an analysis to be worthwhile since ultimately the actinides must be produced
 in an r-process environment.  Indeed, it is possible to produced the second and third r-process peaks without  producing actinides (cf.~\citet{woosley94}).  Moreover, 
 the actinides are particularly sensitive to the production of seed nuclei by light-element reactions 
\citep{sasaqui05a,sasaqui05b} and are also important for cosmochronology.

Even so, we note that there are additional uncertainties associated with the formation of the
actinide nuclei due for example to uncertainties in atomic mass extrapolations, fission barriers,
beta-delayed fission, etc.  Nevertheless, such an application is within the spirit of the schematic
model analysis applied here and in \citet{hoffman97} and provides additional insight into the plausible 
conditions for a successful r-process.

We adopt the following values in calculating the r-process production: an
initial electron fraction of $Y_{e,i}$, 0.45 and dynamical time-scales from 1- 50 msec. In most successful simulations $Y_{e,i}$ remains fixed at
near 0.45 by the ambient weak interaction rates. Hence, although these results will change for different values of $Y_e$, we adopt a fixed value for this figure.  On the other hand, various values of $\tau_{dyn}$ have been proposed in the literature. %We choose a spherical steady-state flow model for the neutrino-driven %wind of \citet{otsuki03} as a standard, --because this flow model is one %of the current models which lead to a successful %r-process--, and
We then search for entropy values for which the r-process abundance distribution is consistent with observation for each adopted dynamical timescale. Examples of consistent entropy values are summarized in
Figure \ref{f5}. The right most figures
  roughly correspond to expansion timescales studied in \citet{hoffman97}. For $Y_e=0.45$, they
  obtained minimum entropies of 140 (5 msec) and 300 (50 msec). That is
  about a factor of two, and factor of 6 respectively less than the values deduced in the present work.
These results, however, will change for different values of  $Y_e$ as is evident from Eqs.~(\ref{result1}) or (\ref{result2}).
%Because even if the entropy per baryon is moderately low %$s/{\rm k}\approx$ 100-200, that the r-process can occur in this %neutrino-driven wind when the dynamical expansion timescale becomes much %shorter than the collision timescale of neutrino-nucleus interactions. 

Figure \ref{f1-2-3} shows the relation between the analytic model (dotted lines) and
numerical simulation (points).
%In figure \ref{f1-2-3}, in model parameters, %some results are selected.
%is on set $\bar{A}$.
Shown are the lower limits on the entropy required to form A=232 (Th) nuclei consistent with observation. Both results are similar.
Figure \ref{f1-2-3} also shows a comparison between the present lower limits and those of Hoffman et al.~(1997). The new relation implies that the required entropy is typically a factor of two greater than the previous $s/k$-$\tau_{dyn}$ estimate.

In summary, we have shown that the \Li ~reaction is an important competing reaction flow channel for r-process nucleosynthesis. This reaction in particular implies a more efficient production of seed nuclei so that a larger neutron/seed ratio is required for a successful dynamical r-process model. For the schematic exponential models considered
here, the implied lower limit to the entropy per baryon increases by about a factor of two from previous estimates. This places a serious constraint on models for the astrophysical site for the
production of r-process nuclei.

%\section{Summary}
%wedge product ()$^{\alpha}\times$()$^{\beta}$ wo kangaeruka? --> %*\alpha=1, \beta=1 ---> onajikurai, i.e. 50%50% important 

%\subsection{} %\label{bozomath}
\acknowledgments
This work has been supported in part by Grants-in-Aid for Scientific Research (12047233, 13640313, 14540271) and for Specially Promoted Research (13002001) of the Ministry of Education, Science, Sports and Culture of Japan, and The Mitsubishi Foundation. Work at UND supported under DoE nuclear theory grant $\#$DE-FG02-95-ER 40934.

%Facilities: \facility{Nickel}, \facility{HST(STIS)}, \facility{CXO(ASIS)}. 

\appendix

\section{}

\clearpage
\section{Appendix A}
The following updated expressions were utilized to calculate the reaction rates $N_A \langle\sigma
v\rangle$. Although these rates differ from those used in \citet{hoffman97}, employing them does not substantially change the results. 

% -------------------------------------------------------------------- 
For the reaction 
$^{9}$Be($\alpha$,n)$^{12}$C~reaction we use; \begin{eqnarray}
N_A \langle \sigma v\rangle_{\rm{NOW}} &=& 4.62\times 10^{13}/{T_9^{2/3}}
\exp{(-23.870/{T_9^{1/3}}-(T_9/0.049)^2)}\nonumber \\ &&\times(1.+0.017\times{T_9^{1/3}}+8.57\times{T_9^{2/3}} +1.05\times{T_9}+74.51\times{T_9^{4/3}}+23.15\times{T_9^{5/3}})\nonumber \\ &&+7.34\times 10^{-5}/{T_9^{3/2}}\exp(-1.184/{T_9})\nonumber \\ &&+0.227/{T_9^{3/2}}\exp(-1.834/{T_9})\nonumber \\ &&+1.26\times 10^{5}/{T_9^{3/2}}\exp(-4.179/{T_9})\nonumber \\ &&+2.40\times 10^{8}\times \exp(-12.732/{T_9}).\nonumber \end{eqnarray}
On the other hand Hoffman et al.(1997) used Wrean et al.(1994); \begin{eqnarray}
N_A \langle\sigma v\rangle_{{\rm WREAN}} &=& 6.476\times 10^{13}/{T_9^{2/3}}\exp(-23.8702/T_9^{1/3}) \times(1.0-0.3270\times {T_9^{1/3}})\nonumber \\ &&+6.044\times 10^{-3}/{T_9^{3/2}}\exp(-1.041/{T_9})\nonumber \\ &&+7.268/{T_9^{3/2}}\exp(-2.063/{T_9})\nonumber \\ &&+3.256\times 10^{4}/{T_9^{3/2}}\exp(-3.873/{T_9})\nonumber \\ &&+1.946\times 10^{5}/{T_9^{3/2}}\exp(-4.966/{T_9})\nonumber \\ &&+1.838\times 10^{9}/{T_9^{3/2}}\exp(-15.39/{T_9}).\nonumber \end{eqnarray}
We use the newest reaction rate for $^{8}$Li($\alpha$,n)$^{11}$B~ from X.Gu et al.(1995) and Hashimoto et al.(2005); \begin{eqnarray}
N_A \langle\sigma v\rangle_{\rm Gu} &=&
4.929\times 10^{6}/{T_9^{3/2}}\exp(-4.410/{T_9})\nonumber \\ &&+ 5.657\times 10^{8}/{T_9^{3/2}}\exp(-6.846/{T_9})\nonumber \\ &&+ 4.817\times 10^{9}/{T_9^{3/2}}\exp(-11.836/{T_9})\nonumber \\ &&+ 1.0\times 10^{12}/{T_9^{2/3}}\exp(-19.45/{T_9^{1/3}}) \times(10.03/{T_9^{1/3}} + 4.814).\nonumber \end{eqnarray}
Then we can get numerically these results; \begin{eqnarray}
\int_{2.5}^{5.0}f(T_9){\rm d}T_9=\left\{ \begin{array}{ll}
6.4\times 10^8 s^{-3} & (\rm{for~Wrean~et~al.~(Hoffman~et~al.~97)}) \\ 7.4\times 10^8 s^{-3} & (\rm{for~our~version}) \\ \end{array} \right.
\end{eqnarray}
\begin{eqnarray}
\int_{2.5}^{5.0}h(T_9){\rm d}T_9=\left\{ \begin{array}{ll}
1.5\times 10^4 s^{-7/2} & (\rm{for~X.Gu~et~al.~version}) \\ 3.6\times 10^3 s^{-7/2} & (\rm{for~Hashimoto~et~al.~version}) \\ \end{array} \right.
\end{eqnarray}

\clearpage

\section{Appendix B}
Eq.~(\ref{13e}) is solved by the following mathematical method. \\ There exists such a function $y(t)$ that satisfies this differential equation,
\begin{eqnarray}
{{{\rm d}y}\over{{\rm d}t}} = f(t)y+ h(t)y^2 ~~, \end{eqnarray}
where $f(t)$ and $h(t)$ are any functions of $t$. \\ First, we replace \(y^{-1} = z\). Then \(z^{'}=-y^{-2}y^{'}\). This leads to the homogeneous first order differential equation, \begin{eqnarray}
{{{\rm d}z}\over{{\rm d}t}} + zf(t) = -h(t) ~~. \end{eqnarray}
Next, multiplying the both sides by $e^{F(t)}$, here \(F(t)=\int f(t) {\rm d} t\), we have,
\begin{eqnarray}
{{\rm d}\over{{\rm d}t}}\biggl(z(t)e^{F(t)}\biggr) = -h(t)e^{F(t)}. \nonumber \end{eqnarray}
Now integrating the above equation between \(t= [t_i:t_f]\) we have, \begin{eqnarray}
z(t_{f})e^{F(t_{f})}-z(t_{i})e^{F(t_{i})}= -\int_{t_{i}}^{t_{f}}h(t)e^{F(t)}{\rm d}t,\nonumber \end{eqnarray}
\begin{eqnarray}
z(t_{f})=z(t_{i})e^{F(t_{i})-F(t_f)}
-e^{-F(t_{f})}\int_{t_{i}}^{t_{f}}h(t)e^{F(t)}{\rm d}t, \nonumber \end{eqnarray}
where \(F(t_{i})-F(t_f) = \int_{t_{f}}^{t_{i}}f(t){\rm d}t\).

Finally, transforming the valuable $z(t)$ back to $y(t)$, we can get the solution:
\begin{eqnarray}
{1\over{y_{n,f}}}={1\over{y_{n,0}}}\exp{\Biggl[ \int_{t_f}^{t_i}f(t){\rm d}t \Biggr]}
+\int_{t_f}^{t_i}h(t)
\exp{\biggl[\int_{t_f}^{t}f(t^{'}){\rm d}t^{'}}\biggr] {\rm d}t. \label{ge-sol}
\end{eqnarray}

\clearpage
\section{Appendix C}
Let us suppose that there exists such a function $p(T)$ such that $p(T)$ is very large for the $T_{i}$ of interest. Then,
\begin{eqnarray}
p(T) \approx p(T_{i})\exp\Biggl\{-\biggl[{{{\rm d}\ln{p(T)}}\over{{\rm d}T}}\biggr]_{T_{i}} (T_{i}-T)\Biggr\}. \nonumber \end{eqnarray}
Because \(p(T)=\exp[\ln{p(T)}]\) and \(\ln{p(T)} \approx \ln{p(T_{i})} + \biggl[{{{\rm d}\ln{p(T)}}\over{{\rm d}T}}\biggr]_{T_{i}}(T-T_{i})\).\\ \subsection{Application}
We apply the above relation to the 2nd term on the right hand side of
Eq. (15). Thus, we have
\(p(T)=\alpha_1h(T)\exp\biggl[
\alpha_{0}\int_{T_f}^{T}f(T^{'}){\rm d}T^{'}\biggr]\), where \(\alpha_{0} = (\bar{A}-2\bar{Z})\tau_{dyn}Y_{\alpha}^{3}\), \(\alpha_{1} = (\bar{A}-2\bar{Z})\tau_{dyn}Y_{\alpha}^{3/2}\), \(T_i=5.0\), and \(T_f =2.5\).
Clearly, $p(T)$
satisfies the condition that $p(T=5.0)$ is large. Since,
\begin{eqnarray}
p(T_i)=\alpha_1h(T_i)\exp\biggl[
\alpha_{0}\int_{T_f}^{T_i}f(T^{'}){\rm d}T^{'}\biggr],\nonumber\\ \ln{p(T)}=\ln{\alpha_1h(T)}+\alpha_{0}\int_{T_f}^{T}f(T^{'}){\rm d}T^{'},\nonumber\\ {{{\rm d}\ln{p(T)}}\over{{\rm d}T}}={{h^{'}(T)}\over{h(T)}}+ \alpha_0 f(T),\nonumber
\end{eqnarray}
we can deduce the following formula:
\begin{eqnarray}
p(T)\approx \biggl\{\alpha_1h(T_i)e^{
\alpha_{0}\int_{T_f}^{T_i}f(T^{'}){\rm d}T^{'}}\biggr\} \times
\exp\biggl[-\biggl({{h^{'}(T_i)}\over{h(T_i)}}+ \alpha_0 f(T_i)\biggr) (T_i-T)\biggr] ~~, \end{eqnarray}
where we have replaced \(\biggl({{h^{'}(T_i)}\over{h(T_i)}}+ \alpha_0 f(T_i)\biggr)\) with $\beta_0$. \begin{eqnarray}
\int_{T_f}^{T_i}p(T){\rm d}T &\approx&
\biggl\{ \alpha_1h(T_i)e^{
\alpha_{0}\int_{T_f}^{T_i}f(T^{'}){\rm d}T^{'}}\biggr\} \times
\int_{T_f}^{T_i}e^{-\beta_0 (T_i-T)}{\rm d}T\nonumber \\ &=&
\biggl\{ \alpha_1h(T_i)e^{
\alpha_{0}\int_{T_f}^{T_i}f(T^{'}){\rm d}T^{'}}\biggr\} \times
{{1}\over{\beta_0}}\biggl[1-e^{-\beta_0(T_i-T_f)}\biggr].\nonumber \end{eqnarray}

\clearpage

\clearpage

\clearpage

%\begin{figure}
%\centering
%\rotatebox{0}{\includegraphics[width=18cm,height=15cm]{f4_ver4.pdf}} 
%\rotatebox{0}{\includegraphics[width=18cm,height=15cm]{f3.eps}} 
%\caption{The minimum entropy required
%to produce the observed abundance of r-process nuclei based upon the numerical
%network calculation (points and dot-dashed lines) and
%the analytic model (dotted lines) as a function of the dynamical timescale.
%The solid lines show the result of considering only the \aan flow % of Eq.~(\ref{result1}) ($i.e.$ \citet{hoffman97}) or the % \atg (n,$\gamma$)\Li ~flow alone as labeled. %The comparison of the model result, calculation one and the 
%hitherto calculation of \citet{hoffman97}. % The points are the calculation result of entropy 
%for each dynamical timescale, each value are suitable for the % observation (Figure \ref{f5}).
%The thick solid line is the new entropy lower limit for 
%successful actinide production, the thick dashed line is the minimum entropy required to reproduce the 3rd r-process peak, while the thick dotted line is for the 2nd peak. The thin lines show the analogous constraints deduced from Eq.~(\ref{result1}) ($i.e.$ \citet{hoffman97}) 
%based upon the flow through the \aan ~reaction alone.} \label{f4}
%\end{figure}

\clearpage

\end{document}